# Early Compton Effect experiments revisited; Evidence for outstanding hard X- ray reflectivity of speculum metal

Nazimudeen E. A[a], Girish T. E [b*], Sunila Abraham[a]

[a] Department of Physics, Christian College, Chengannur, Kerala, India – 689122

[b] Department of Physics, University College, Thiruvananthapuram, Kerala, India – 695034

*Author for correspondence: tegirish5@yahoo.co.in

**Abstract**

Experiments related to X- ray optics carried out by A. H. Compton and his collaborators during 1923 – 1927, led to the Nobel Prize winning discovery of Compton Effect. Based on the critical analyses of these experimental invetigation results, we could infer a unique X- ray reflection property of speculum metal that provided us the evidence for its reflectivity in the hard X-ray region. In connection with this, we have studied in detail the theoretical X-ray reflectivity of speculum metal in comparison with variety of other mirror materials for different photon energies at various grazing angles of incidence. Theoretical calculations of grazing incidence X-ray reflectivity for these materials also verified well with other experimental data and computer simulations. We have also calculated the minimum and critical angle penetration depth of speculum metal with other mirror materials for different photon energies. Furthermore, the reflectivity of speculum metal in different regions of electromagnetic spectrum in association with its physical properties is discussed. The analytical results obtained from these studies indicated that the reflectivity of speculum metal in the X- ray region, especially in hard X-ray region at small grazing angles is higher than that for commonly used high density mirror materials.

## 1. Introduction

The total reflection of X- rays from solid samples with smooth surfaces first reported by A. H. Compton in 1923 can be assumed to mark the birth of grazing incidence X-ray optics [1- 4]. With this technique the field of X- ray astronomy became a major scientific discipline in astrophysics and cosmology [5, 6]. X-ray astronomy has made great progress in the past several years with technological advances in X-ray observations and instrumentations. The remarkable scientific observations in X-ray astronomy have enriched and guided new views of the universe. They have shed light upon dark matter and dark energy [7, 8]. However, future space missions in X-ray astronomy requires the integration of highly reflective thin mirror segments into a light weight, high resolution telescope housing without distortion [9]. Mirror materials like nickel, gold, platinum and Iridium are good reflectors and are commonly used for making X-ray mirrors in space missions [10]. But recent advances in grazing incidence X-ray optics, especially the advent of synchrotron radiation sources have stimulated the need for basic research in high quality optically active mirror materials for novel applications. Fortunately, we could infer a brilliant X-ray reflection property of speculum metal from early experiments related to X-ray optics carried out by A. H. Compton and R. L. Doan, that provided us the evidence for its high reflectivity in the hard X-ray region [11, 12]. In this context a detailed study of the reflectivity of speculum metal in different regions of electromagnetic spectrum, especially in hard X- ray region will be highly useful, that forms the theme of the present paper. Thus we have studied theoretically the X- ray reflection properties of various mirror materials including speculum metal as functions of different photon energies and various grazing angles of incidence.

Speculum metal is a typical binary alloy of copper and tin in the proportion of about 2:1, whose peculiar properties lies in its excellent resistance to tarnishing, pleasing silvery white color and brilliant polishing. The antiquity of this particular alloy dates back to the Bronze Age [13- 15]. It regained its prominence for a short period from the time of Isaac Newton in 17th century and continuing up to early decades of 20th century as mirrors, gratings, etc. for making large reflecting telescopes and other optical precession instruments [16, 17]. Speculum metal is not a material of the past, but it is still relevant today because of its excellent applications in diverse fields of science and technology apart from its archaeometallurgical importance. In recent years it finds several engineering applications in various fields such as slit material in transmission electron microscopy, alternative to nickel undercoating for gold or chromium electroplating, anode material for lithium-ion batteries, etc. [19- 20]. It can also be used in several scientific observations and discoveries covering different areas including experimental astronomy, optical and nuclear instrumentations, spectroscopic and other studies in modern physics [21- 23].

## 2. Basic concepts of X-ray reflection

The interaction of X- rays with matter can be described by an index of refraction less than unity, and is given by,

$$n = 1 - \delta - i\beta \qquad (1)$$

With real part $\delta$, a decrement of the refractive index describing the reflection, given by

$$\delta = \frac{N_A}{2\pi} r_e \frac{Z}{A} \rho \lambda^2 = C_m \lambda^2 \qquad (2)$$

and $\beta$ determining the absorption properties of the material, given by $\beta = \frac{\lambda \mu}{4\pi}$, where $N_A$ is the Avogadro number, $r_e$ is the classical electron radius, Z is the atomic number, A is the atomic mass and $\rho$ is the mass density, $\lambda$ is the wavelength of X-ray photon, $C_m$ is the material constant and $\mu$ is the linear absorption coefficient. X- rays undergo total external reflection only for grazing angle of incidence ($\theta$) less than the critical angle,

$$\theta_c = (2\delta)^{1/2} = \frac{1.651}{E} \left(\frac{Z}{A}\rho\right)^{1/2} \qquad (3)$$

Where E is the photon energy. The depth of penetration is calculated by following approximation relations,

(i) $\quad Z = \frac{\lambda}{4\pi} \frac{1}{\sqrt{2\delta}} \qquad$ when $\theta < \theta_c \qquad$ (4. a)

(ii) $\quad Z = \frac{\lambda}{4\pi} \frac{1}{\sqrt{\beta}} \qquad$ when $\theta = \theta_c \qquad$ (4. b) and

(iii) $\quad Z = \frac{\lambda}{4\pi} \frac{\theta}{\beta} \qquad$ when $\theta > \theta_c \qquad$ (4. c)

Reflection of X- rays is described by the Fresnel equations, which are derived by Compton and Allison. Reflectivity is calculated by three approximation relations, given by

(i) $\quad R = 1 - \sqrt{\frac{2}{\delta} \cdot \frac{\beta}{\delta}} \, \theta \quad$ when $\theta = \theta_c \qquad$ (5. a)

(ii) $\quad R = \frac{\delta + \beta - \sqrt{2\beta\delta}}{\delta + \beta + \sqrt{2\beta\delta}} \quad$ when $\theta < \theta_c \qquad$ (5. b) and

(iii) $\quad R = \frac{\delta^2}{4\theta^4} \qquad$ when $\theta > \theta_c \qquad$ (5. c)

Where $\theta$ is the grazing angle of incidence [24- 30].

## 2. Materials and methods

Experimental investigations are carried out by using a highly reflecting speculum metal mirror, cast and polished using traditional techniques from Aranmula, a village in Kerala, Southern India. There exist literary articles that describe its metallurgical and technical aspects of casting and polishing methods [31- 33]. The purchased mirror (Aditi Handicrafts Centre, Aranmula) with an aperture of 0.05m, separated from the brass base is cleaned and sectioned into number of fragments of different dimensions using diamond cutter, which are used for experimental characterization.

The reflectivity of speculum metal from near infrared to extreme ultraviolet regions of electromagnetic spectrum is studied by compiling the data of the results of previous literary work and our experimental investigations. X-ray reflection parameters are calculated theoretically by using basic equations and verifications of these calculations are carried out by using computer simulation programs such as XCOM, a Standard Reference Data Program of NIST, RefleX, a program based on Henke- Gullikson-Davis database in Lawrence Berkeley Laboratory and Sergey's X-Ray library database in Argonne National Laboratory [34- 36].

## 3. Results and Discussion

3.1. Physico – chemical properties of speculum metal

The chemical composition, surface morphology, structural evaluation and optical reflectance properties at the reflecting surface of cast speculum metal mirror samples are studied in detail. Previous experimental investigation results of the material science characteristics of speculum metal [37- 39] are summarized in Table 1. The detailed chemical composition studies give us a well insight that the alloying system is based on two major elements including copper and tin. The three dimensional AFM morphology at 2 µm shows a streak or texture pattern with porous structure whose RMS surface roughness is about 3.38 nm. It consists of many fine nano grains and each grain consists of elongated rulings with crystals arranged in uniform patterns. XRD pattern consists of different intermetallic phases of Cu- Sn alloy system, in which particle size varied between 29.4 to 80.7 nm. Combined use of XRD and AFM analyses at the polished surface of cast speculum metal suggest that on the surface there exist a thin, transparent layer consisting of varying particles that are dimensionally nanoscale and crystalline in nature.

3.2. Reflectivity of Speculum Metal across Electromagnetic spectrum

We have studied the reflectivity of speculum metal in different regions of electromagnetic spectrum namely near infrared, visible, near ultraviolet, vacuum ultraviolet (Schumann), extreme ultraviolet and X-ray by using data compilations, theoretical calculations and computer simulations . The reflectivity of speculum metal in different regions of electromagnetic spectrum is also compared with other metals.

3.2.1. Reflectivity of Speculum Metal from Near Infrared to Extreme Ultraviolet Region.

The optical reflectivity of speculum metal is measured in a region of wave length ranging from 200 nm to 900 nm using JASCO - 550V UV- Visible Double Beam Spectrophotometer. Polished speculum metal shows a relatively high reflecting power in the near infrared region followed by a uniform and generally, a low reflectivity in the visible and near ultraviolet spectrum. In the lower wavelength region, there is an absorption, which results in a significant loss of energy. The spectral reflecting power data indicate that in the ultraviolet spectrum the reflectivity is only 37 per cent at 350 nm, which gradually increases to 67 per cent at 650 nm in the visible spectrum and 74 per cent at about 900 nm in near infrared. These results are in good agreement with the result of previous experimental investigations [32, 40], which are reported in **Table 2**. Reflectivity of Speculum metal in the vacuum (Schumann) and extreme ultraviolet regions are studied by compiling the data of the results of previous experimental investigations, which shows poor reflection approximately 7.5 % in the region of extreme ultraviolet and 10 – 12% in the vacuum ultraviolet region [41, 42]. According to Gardner [43] the reflecting power of copper, nickel and gold show only minor variations of reflection when compared with speculum in Schumann region, but silver and aluminium are much poorer than speculum metal in that region.

### 3.2.2. Reflectivity in the X- ray region

### 3.2.2.1. Experimental Evidence for High X-ray Reflectivity of Speculum Metal

We have studied in detail the reflectivity of speculum metal in the X-ray region based on the Nobel Lecture report of A. H. Compton and some early literary work [2, 44]. Compton pointed out in his literary article that considerable X-ray reflection does occur from a speculum surface when it strikes the surface at small glancing angles, within the critical angle for total reflection [15, 16]. Latterly, Doan confirmed this fascinating experimental investigation result in collaborated with Compton and they obtained the spectra of ordinary X-rays by reflection at very small glancing angles from a grating ruled on speculum metal [23]. Some of these spectra are used to measure X-ray wave-lengths with considerable precision. Spectra of K – series radiations both from copper and molybdenum obtained by reflection at a small glancing angle from a grating ruled on speculum metal are shown in Fig. 1 (a) and (b).

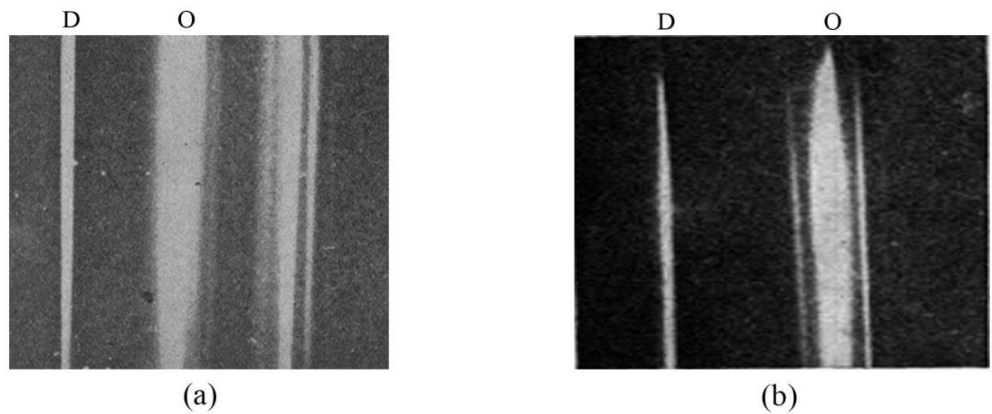

Fig. 1. (a) Reflection on speculum using Cu $K_a$ X- rays [23]. D – Image of direct beam, O - Directly reflected beam.
(b) Reflection on speculum using Mo $K_a$ X- rays [15, 23, 37]

3.2.2.2. Theoretical Calculations based on X-ray Reflection.

Based on the fundamental theory of X- ray reflection, first we have calculated the total external reflection critical angle for speculum metal with other mirror materials for photon energies from 8.05 keV to 300 keV, which are given in table 2 and graphically sown in Fig. 2. It is observed from our study that the critical angle reduces as the x-ray photon energy increases from 8 keV to 300 keV and also observed a higher value critical angle for iridium and low for aluminium. These observations led us to conclude that the critical angle for total external reflection reduces inversely proportional to the photon energy and also high density materials with higher atomic number have larger critical angle at any photon energy. Early literary works proposed by N. V. Alov and Klockenkämper reported the total reflection critical angle of Al, Ni, Cu, Pt and Au, which are very close to our estimation [25,45 ].

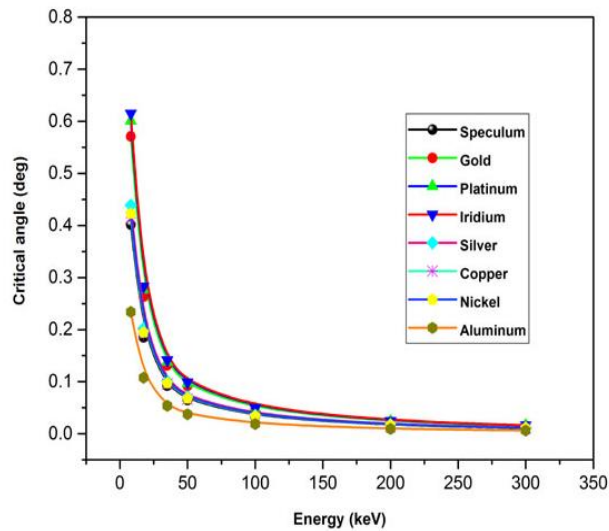

Fig. 2. Critical angle for various materials calculated for different photon energies

Further, we have calculated theoretically the minimal and critical angle penetration depth of speculum and other mirror materials as functions of different photon energies, which are displayed in Table 3 and also graphically shown in Fig. 3. It is noticed from our calculations that the minimal and critical angle penetration depth are high for aluminium and low for iridium. Minimum penetration depth depends neither on density of the material nor energy of photon, but it changes by varying the number of electrons per unit volume or atomic number. It is also observed that critical angle penetration depth of speculum metal in comparison with other materials increases on a regular basis as the photon energy going on increasing.

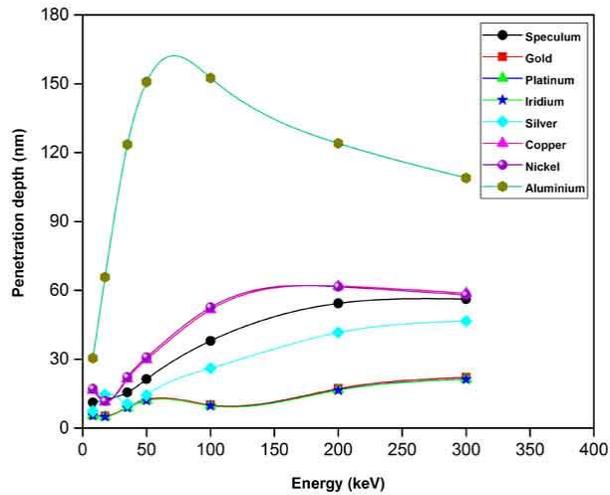

**Fig. 3. Critical angle penetration depth of various materials as functions of different photon energies.**

In the next step, we have studied theoretically the critical angle reflectivity of speculum and other materials for photon energies from 8.05 keV to 300 keV. Critical angle reflectivity of various materials calculated for X-rays of different photon energies are graphically pictured in Fig. 4. Critical angle reflectivity of speculum is found to be gradually increased from 54.2 to 88.3 for photon energies from 8 keV to 300 KeV, compared with other materials. It is also found that critical angle reflectivity of speculum is higher than that of high density common x-ray mirror materials such as iridium, platinum, gold and comparable with that of lighter elements such as silver, nickel and copper. Critical angle reflectivity of speculum with other materials for different photon energies are given in table 4, which are in good agreement with the value reported by early literary works [25, 45].

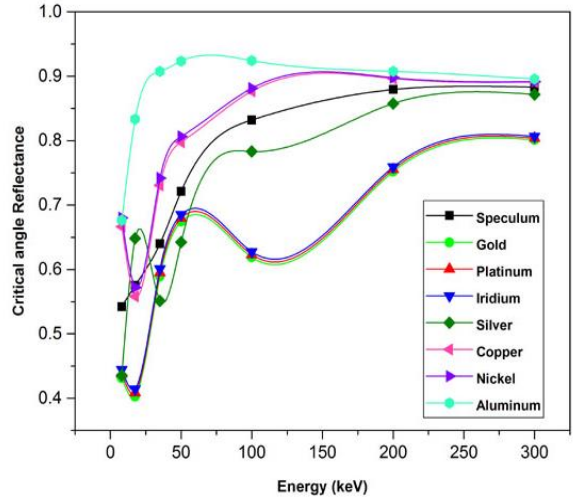

Fig. 4. Critical angle reflectivity of various materials as a function of different photon energies

Based on theoretical calculations and simulations, we have studied the reflectivity of polished speculum metal with other materials at various grazing angles of 1- 10 arc minutes for Cu Kα and Mo K$_α$ radiations with energy 8.05 keV and 17.5 keV, which are compared with the data of other materials obtained from previous literary works [27, 39, 47]. In the soft X-ray region speculum shows a higher reflectivity approximately 96%, which is comparable with that of commonly used high density X-ray mirror materials, but slightly less than that of lighter elements. The reflectivity in the hard X-ray region is found to be about 93% , which is higher than that of commonly used high density X-ray mirror materials and comparable that of light elements. But aluminium shows poor reflection in this hard X-ray region, which is about 1% only. The spectral reflecting power of speculum metal in different regions of electromagnetic spectrum, especially in hard X-ray region compared with other mirror materials are shown in Figs. 5 and 6.

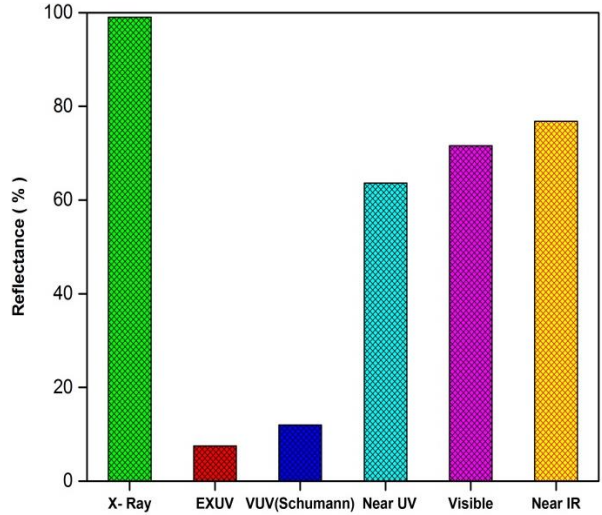

**Fig. 5.** Reflectivity of speculum metal in different regions of electromagnetic spectrum

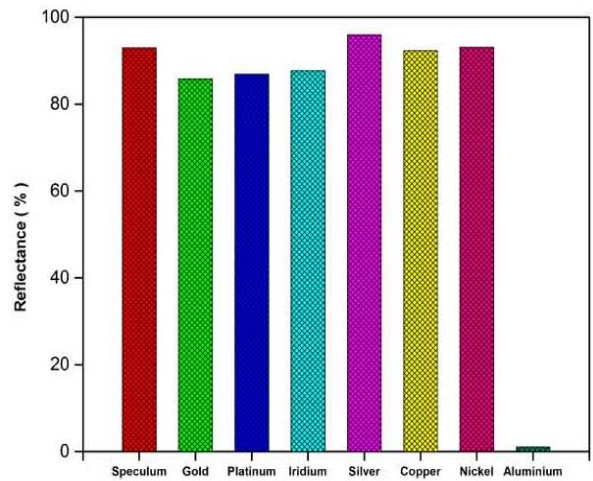

**Fig. 6.** Reflectivity of speculum metal compared with other mirror materials in the hard X-ray region (Mo K$_\alpha$ line)

Recent literary work by Henke et al. also reported the reflectivity of Al, Ni, Cu and Pt at various grazing angles for different energies, which has close resemblance to our estimation [44]. It also suggested that the reflectivity of gold, copper and nickel in the soft X- Ray region (using Copper Kα, particularly for energy 8 keV at a grazing angle of 5 mrad) are above 88 to 96%, but aluminium shows a reflectivity of 7.6%. However in the hard X- Ray region (using Molybdenum Kα with energy 17.479 keV at a grazing angle of 5 mrad) all these materials show poor reflectivity.

Theoretical reflectivity of speculum metal as functions of different photon energies and various grazing angles are graphically shown in fig. 6.

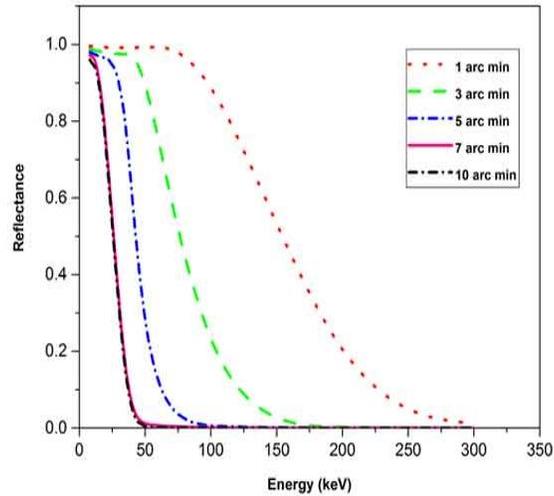

**Fig. 6.** Reflectivity of speculum metal calculated for Mo Kα X-rays at different grazing angles

The reflectivity plot shows that speculum reflects with high efficiency at energies 8.05 keV and 17.5 keV and it decreases very rapidly at higher photon energies at a grazing angle of 10 arc minute. However, it reflects with high efficiency and gives a finite value up to 300 keV at a grazing angle of one arc minute. This implies that the efficiency of reflection is high as the grazing angle within the critical angle regime reduces to its lower value. In brief, the reflectivity of speculum metal in the hard X- ray region is particularly interesting in comparison with other metals and it may open up promising possibilities for this alloy as mirrors, gratings, filters and light pipes, etc. in various fields of astrophysics, especially in X-ray astronomy, X-ray microscopy, X-ray imaging including medicine and industry, X-ray lithography, X-ray reflection spectroscopy and other areas of high energy physics.

## 4. Conclusions

The reflectivity of speculum metal from near infrared to X-ray regions of electromagnetic spectrum has been studied in detail in association with its physical properties, which suggests poor reflectivity in the extreme or vacuum ultraviolet region. X-ray reflectivity of various materials including speculum metal is studied in detail for different photon energies at various grazing angles of incidence. Speculum metal shows a gradual increase in critical angle penetration depth and critical angle reflectivity as the energy of photon increased from 8 keV to 300 keV. Low density materials generally show high reflectivity at small grazing angles compared to high density materials. The reflectivity of speculum metal in the hard X-ray region (Mo Kα) at various grazing angles of 1 – 10 arc minutes is found to be high in comparison with commonly used high density mirror materials such as gold, platinum and iridium. The high reflectivity may be due to the presence of a thin transparent layer of well-polished smooth surface containing nanoscale crystalline particles as evident from XRD and AFM analyses. The high reflectivity of speculum metal in the hard X-ray region will find excellent applications for this typical alloy as mirrors, gratings, filters and light pipes, etc. in various fields of astrophysics, especially in X-ray astronomy, X-ray optics, X-ray reflection spectroscopy and other areas of high energy physics.

**Acknowledgements**

We are extremely grateful to Dr. S. NarayanaKalkura, Dr. K. MuraleedharaVarier, Mr. Midhun, Dr. Gopchandran, Dr. Abdul Kalam, for their constructive comments, valuable suggestions, continuing support and contributions during the entire work, the director of NCESS, the staffs of SICC, University of Kerala, for providing useful reference materials and laboratory facilities to conduct experimental analyses.

**Figure captions**

Figs. 1 (a).   Reflection on speculum using Cu $K_\alpha$ X- rays

   (b).   Reflection using Mo $K_\alpha$ X- rays

Fig. 2.   Critical angle for various materials calculated for different photon energies

Fig. 3.   Critical angle reflectivity of various materials as a function of different photon energies

Fig. 4.   Critical angle penetration depth of various materials as functions of different photon energies

Fig. 5.   Reflectivity of speculum metal across electromagnetic spectrum

Fig. 6.   Reflectivity of speculum compared with other materials in the hard X-ray region (Mo $K_\alpha$ line)

Fig. 7.   Reflectivity of speculum metal calculated for Mo Kα X-rays at different grazing angles

**Tables**

Table1.   Summary of Previous experimental investigation results of the material science characteristics of speculum metal

Table 2.   Reflecting power of various speculum metals in UV region

Table 3.  Critical grazing angle for various materials for X-rays of different photon energies

Table 4.  Minimum and critical angle penetration depth for various materials for X-rays of different photon energies

Table 5.  Critical angle reflectivity of various materials for X-rays of different photon energies

Table 6.  Reflectivity of various materials calculated for Mo K$_\alpha$ X-rays at different grazing angles

**Table 1.  Summary of Previous experimental investigation results of the material science characteristics of speculum metal**

| Experimental Technique | Analyzed Parameter | Sample in cast form |
|---|---|---|
| EDS | Chemical composition | Cu - 67.3, Sn -30.8, Oths- 1.9 |
| AFM | Surface Roughness | 3.38 nm |
| XRD | Particle Size | 29.4 – 80.7 nm |
|  | Intermetallic phase | $Cu_3Sn$, $Cu_{41}Sn_{11}$, $Cu_{5.6}Sn$, $Cu_{4.8}Sn_4$ |
| UV- Visible Spectrophotometer | Visible reflectance | 67 % |

**Table 2.  Reflecting power of various speculum metals in UV region**

| Wavelength (nm) | Reflecting power of different speculum (Wt. %) | | |
|---|---|---|---|
|  | **Hagen and Rubens** | **Hulbert** | **Aranmula speculum** |
| **251** | 29.9 | 37 | 35.6 |
| **288** | 37.7 | 41 | 43.7 |
| **305** | 41.7 | 44 | 47.2 |
| **357** | 51 | 60 | 57.7 |

**Table 3.** Critical grazing angle for various materials for X-rays of different photon energies.

| Materials | Density ($gcm^{-3}$) | Critical grazing angle ($\theta_C$) in degree for different materials for X-rays of different photon energies. | | | | | | |
|---|---|---|---|---|---|---|---|---|
| | | 8.05 keV | 17.5 keV | 35 keV | 50 keV | 100 keV | 200 keV | 300 keV |
| **Speculum** | 8.62 | 0.4017 | 0.1849 | 0.0923 | 0.0646 | 0.0323 | 0.0161 | 0.0107 |
| **Gold** | 19.31 | 0.5709 | 0.2628 | 0.1312 | 0.0918 | 0.0459 | 0.0229 | 0.0153 |
| **Platinum** | 21.45 | 0.6007 | 0.2766 | 0.1381 | 0.0967 | 0.0483 | 0.0241 | 0.0161 |
| **Iridium** | 22.4217 | 0.6148 | 0.2830 | 0.1413 | 0.0989 | 0.0494 | 0.0247 | 0.0164 |
| **Silver** | 10.5 | 0.4388 | 0.2020 | 0.1008 | 0.0706 | 0.0353 | 0.0176 | 0.0117 |
| **Copper** | 8.96 | 0.4148 | 0.1909 | 0.0953 | 0.0667 | 0.0333 | 0.0166 | 0.0111 |
| **Nickel** | 8.90225 | 0.4227 | 0.1946 | 0.0972 | 0.0680 | 0.0340 | 0.0170 | 0.0113 |
| **Aluminum** | 2.6989 | 0.2339 | 0.1077 | 0.0537 | 0.0376 | 0.0188 | 0.0094 | 0.0062 |

**Table 4.** Minimal and Critical angle penetration depth of various materials for different photon energies

| Materials | Minimal depth (nm) | Critical angle penetration depth for various materials at different photon energies (nm) | | | | | | |
|---|---|---|---|---|---|---|---|---|
| | | 8.05 keV | 17.5 keV | 35 keV | 50 keV | 100 keV | 200 keV | 300 keV |
| **Speculum** | 1.75 | 11.23 | 12.49 | 15.53 | 21.32 | 38.00 | 54.35 | 56.21 |
| **Gold** | 1.23 | 5.66 | 5.20 | 9.18 | 12.41 | 10.15 | 17.22 | 22.18 |
| **Platinum** | 1.17 | 5.48 | 5.03 | 8.90 | 12.04 | 9.79 | 16.63 | 21.40 |
| **Iridium** | 1.14 | 5.47 | 5.00 | 8.87 | 12.02 | 9.72 | 16.52 | 21.23 |
| **Silver** | 1.60 | 7.46 | 14.66 | 10.58 | 14.34 | 26.13 | 41.66 | 46.68 |
| **Copper** | 1.69 | 16.57 | 11.45 | 21.49 | 29.89 | 51.58 | 61.91 | 58.77 |
| **Nickel** | 1.66 | 17.13 | 11.75 | 22.18 | 30.85 | 52.63 | 61.58 | 57.99 |
| **Aluminum** | 3.00 | 30.53 | 65.79 | 123.59 | 150.85 | 152.53 | 124.07 | 109.00 |

**Table 5.  Critical angle reflectivity of various materials for X-rays of different photon energies.**

| Energy (keV) | Critical angle reflectivity ($R_C$) of various materials for X-rays of different photon energies. | | | | | | | |
|---|---|---|---|---|---|---|---|---|
| | Speculum | Gold | Platinum | Iridium | Silver | Copper | Nickel | Aluminum |
| 8.05 | 0.5421 | 0.4313 | 0.4379 | 0.4446 | 0.4357 | 0.6663 | 0.6801 | 0.6764 |
| 17.5 | 0.5753 | 0.4026 | 0.4086 | 0.4144 | 0.6484 | 0.5582 | 0.5724 | 0.8333 |
| 35 | 0.6397 | 0.5889 | 0.5949 | 0.6010 | 0.5511 | 0.7306 | 0.7419 | 0.9074 |
| 50 | 0.7213 | 0.6744 | 0.6798 | 0.6851 | 0.6421 | 0.7975 | 0.8065 | 0.9234 |
| 100 | 0.8321 | 0.6188 | 0.6231 | 0.6274 | 0.7831 | 0.8770 | 0.8814 | 0.9242 |
| 200 | 0.8793 | 0.7520 | 0.7555 | 0.7590 | 0.8576 | 0.8964 | 0.8977 | 0.9077 |
| 300 | 0.8830 | 0.8013 | 0.8040 | 0.8066 | 0.8718 | 0.8911 | 0.8917 | 0.8957 |

**Table 6.  Reflectivity of various materials calculated for Mo $K_\alpha$ X-rays at different grazing angles.**

| Material | Reflectivity of various materials calculated for Mo $K_\alpha$ X-rays at different grazing angles | | | | |
|---|---|---|---|---|---|
| | 1 arc min | 3 arc min | 5 arc min | 7 arc min | 10 arc min |
| Speculum | 99.29 | 97.89 | 96.48 | 95.04 | 92.92 |
| Gold | 98.58 | 95.74 | 92.90 | 90.02 | 85.77 |
| Platinum | 98.70 | 96.10 | 93.49 | 90.85 | 86.95 |
| Iridium | 98.77 | 96.32 | 93.86 | 91.37 | 87.68 |
| Silver | 99.61 | 98.82 | 98.04 | 97.24 | 96.06 |
| Copper | 99.24 | 97.71 | 96.18 | 94.64 | 92.35 |
| Nickel | 99.31 | 97.95 | 96.59 | 95.20 | 93.15 |
| Aluminum | 99.87 | 99.61 | 99.36 | 4.50 | 1.09 |